\documentstyle[prl,aps,amssymb]{revtex}
\input epsf.tex

\newcommand{\nc}{\newcommand}
\nc{\beq}{\begin{equation}}
\nc{\eeq}{\end{equation}}
\nc{\beqa}{\begin{eqnarray}}
\nc{\eeqa}{\end{eqnarray}}
\nc{\lra}{\leftrightarrow}
\def\sfrac#1#2{{\textstyle\frac#1#2}}
\nc{\sss}{\scriptscriptstyle}
{\nc{\lsim}{\mbox{\raisebox{-.6ex}{~$\stackrel{<}{\sim}$~}}}
{\nc{\gsim}{\mbox{\raisebox{-.6ex}{~$\stackrel{>}{\sim}$~}}}

\newcommand{\da}{\dot{a}}
\newcommand{\db}{\dot{b}}

\newcommand{\dda}{\ddot{a}}
\newcommand{\ddb}{\ddot{b}}

\newcommand{\pa}{a^{\prime}}
\newcommand{\pb}{b^{\prime}}
\newcommand{\pn}{n^{\prime}}
\newcommand{\ppa}{a^{\prime \prime}}

\newcommand{\ppn}{n^{\prime \prime}}
\newcommand{\fda}{\frac{\da}{a}}
\newcommand{\fdao}{\frac{\dot a_0}{a_0}}
\newcommand{\fdb}{\frac{\db}{b}}

\newcommand{\fdda}{\frac{\dda}{a}}
\newcommand{\fddao}{\frac{\ddot a_0}{a_0}}
\newcommand{\fddb}{\frac{\ddb}{b}}

\newcommand{\fpa}{\frac{\pa}{a}}
\newcommand{\fpb}{\frac{\pb}{b}}
\newcommand{\fpn}{\frac{\pn}{n}}
\newcommand{\fppa}{\frac{\ppa}{a}}

\newcommand{\fppn}{\frac{\ppn}{n}}
\newcommand{\cN}{{\cal N}}
\newcommand{\rpm}{\rho_{\sss\pm}}

\begin{document}
\twocolumn[\hsize\textwidth\columnwidth\hsize\csname@twocolumnfalse%
\endcsname

\draft

\title{Cosmological Expansion in the Presence of an Extra Dimension}

\author{James M.~Cline $^{a,b}$, Christophe Grojean 
$^{b}$ and G\'eraldine Servant $^{a,b}$ }

\address{$^a$ Physics Department, McGill University,
3600 University Street, Montr\'eal, Qu\'ebec, Canada H3A 2T8;\\
$^b$ Service de Physique Th\'eorique, C.E.A.--SACLAY, F--91191 
Gif-sur-Yvette, France}

\maketitle

\begin{abstract} It has recently been pointed out that global solutions of
Einstein's equations for a 3-brane universe embedded in 4 spatial
dimensions give rise to a Friedmann equation of the form $H \propto \rho$
on the brane, instead of the usual $H \propto \sqrt{\rho}$, which is
inconsistent with cosmological observations.  We remedy this problem by
adding cosmological constants to the brane and the bulk, as in the recent
scenario of Randall and Sundrum.  Our observation allows for normal
expansion during nucleosynthesis, but faster than normal
expansion in the very early universe, which could be helpful for 
electroweak baryogenesis, for example. 

\end{abstract}

\pacs{PACS: 98.80.Cq \hfill McGill 99-25 and Saclay t99/065}
]

During the past year, much has been written about the possibility of
having compactified extra dimensions with large radii \cite{ADD}.  In the
original proposal, $M_P$ was related to the radius $b_0$ of the $N$
compact dimensions by $M_P^2 = M^{2}(M b_0)^N$, where $M$ is the new
fundamental quantum gravity scale, which could in principle be as low as 1
TeV.  If so, this would be a partial solution of the hierarchy problem,
{\it i.e.,} why the weak scale, $M_W$, is 17 orders of magnitude smaller
than the Planck scale, $M_P$: it is because $b_0$ is, for some reason,
much larger than $M^{-1}$.  If $b_0 \gg M_P^{-1}$, as is necessary if $M \sim
M_W$, the particles and fields of the standard model must be restricted to
stay on a 3-dimensional slice (brane) of the full $N+3$ spatial
dimensions; otherwise particle propagation in the new dimensions would
already have been seen in accelerator experiments.  But even with the
restriction of the brane, the idea implies many possibly observable
effects at accelerators.  It also poses severe challenges for cosmology.
In this letter we will address one of the cosmological problems, and 
comment upon an unexpected connection to the question of precisely how
the hierarchy problem is solved using the extra dimensions. 

Our starting point is the observation recently made by Bin\'etruy,
Deffayet and Langlois \cite{BDL,others}
that the Friedmann equation for the
Hubble expansion rate of our 3D universe is modified, even at very low
temperatures, by the presence of an extra dimension, $y$, compactified on a
circle or an orbifold.  Allowing for the possibility of a cosmological
constant $\Lambda_b$ in the full 4 spatial dimensions, called the bulk, the
new Friedmann equation for the scale factor $a$ of our brane is
\cite{BDL,Kaloper}
\beq
\label{BDLeq}
	H^2 = \left({\dot a\over a}\right)^2 
	= \left({\rho_{t}\over 6 M^3}\right)^2 +
	{\Lambda_b\over 6 M^3 },
\eeq
instead of the usual relation, $H = \sqrt{\rho_{t}/3M_P^2}$.
$\rho_t$ is the total (vacuum plus matter) energy density on the brane.
This expression is derived, as will be explained below, from the 5D action
\beq
\label{5daction}
	{\cal S} = \int d^{\,4}x\, dy\, \sqrt{|g|}  \left( 
	\sfrac12 M^3{\cal R}
	 -\Lambda_b + {\cal L}_{\mbox{\tiny brane}} \right) 
\eeq
where the action, ${\cal L}_{\mbox{\tiny brane}}$, for the matter
living on the brane results in a stress-energy tensor parametrized as
$T^\mu_\nu = \delta (by)\,  \mbox{diag} (-\rho_t,p_t,p_t,p_t,0)$.  
An interesting aspect of this result is that fact that,
in order to find consistent
global solutions to the Einstein equations in the $(4+1)$-D spacetime,
it is necessary to add a second brane \cite{foot1}, a mirror of our
own, having equal and opposite energy density.  This topology can be
motivated from string theory.  In
the Ho{\v r}ava--Witten picture \cite{HW} of the nonperturbative
regime of the $E_8\times E_8$ string theory, the string coupling is
interpreted  as an eleventh compact dimension with a
${\mathbb{Z}}_2$ symmetry that truncates the spectrum in order  to
keep only sixteen supercharges in 10D, {\it i.e.,} an ${\cal N}=1$
supersymmetry in 4D after compactification on a Calabi-Yau manifold.
There is good evidence \cite{Mtheory} that over a wide range of energies
the theory behaves like a 5D theory compactified on a
${\mathbb{Z}}_2$ orbifold with two 3-branes, viewed as the remanants of
the 10D hypersurfaces where the $E_8$ gauge groups were living. The
two 3-branes can also be seen as D3-branes of the type {\it I} string
theory \cite{ADD}.

Naively, one would expect that at distances much bigger than the size
of the fifth dimension, the effects of the extra compact dimension
become small corrections to the usual 4D equations; thus when $H^{-1}
\gg b_0$, one should recover the standard cosmology.  However the
presence of the mirror brane contradicts this logic.  Let us choose the
range of the compact coordinate to be $y\in [-1/2,+1/2]$.  The
solutions of the Einstein equations for the scale factor $a(y)$ behave
\cite{BDL,others} like $a_0(1 + A|y|/2)$, with $A\sim \rho_t$.  Because
the points $y=\pm 1/2$ are identified, the derivative is discontinuous
at this point, and $a''/a = A(\delta(y)-\delta(y-1/2))$.  The Einstein
equations identify the delta functions with the energy densities of the
two respective branes.  In the limit as $b_0\to 0$, the two branes
overlap, and their energy densities cancel to first order in $\rho_t$
because they are equal and opposite.  Therefore only terms of order
$(a'/a)^2\sim A^2\sim \rho_t^2$ survive, even at arbitrarily late times
in cosmological history.  The resulting expansion rate (1) is probably
incompatible with big bang nucleosynthesis, which is extremely
sensitive to how the Hubble rate varies with the the energy density,
hence temperature.  Even if one tunes $M$ so that the altered expansion
rate (1) still gives the correct helium, it is likely that the other
elements will come out wrong, since their rates of production depend
quite differently on the temperature.

\vskip 0.5cm
\centerline{\epsfxsize=3.4in\epsfbox{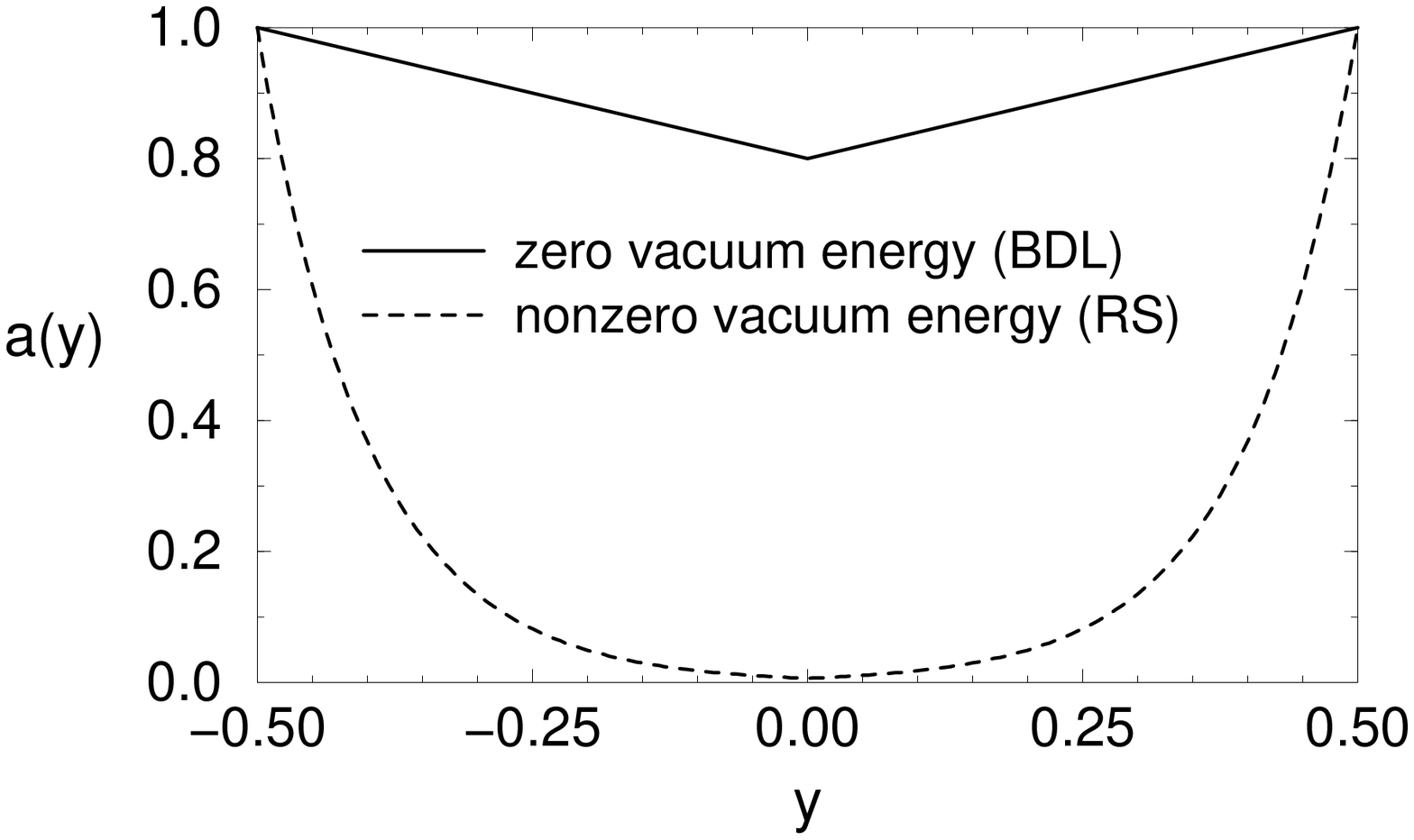}}
\noindent {\small Figure 1. Qualitative dependence of the 3D scale factor 
$a(y)$ on 
the compact dimension $y$ in the solutions of: (solid line) ref.\ \cite{BDL}, 
with vanishing bulk cosmological constant; and (dashed line) ref.\ \cite{RS},
with $\Lambda_b$ given by eq.\ (\ref{Lambda1}). $a(0)$ is nonzero but
exponentially small in the latter.
}
\vskip 0.5cm

From eq.\ (\ref{BDLeq}), one can imagine a very simple escape from this
dilemma \cite{foot2}.  Suppose there is a cosmological constant $\Lambda$
localized on our brane (and correspondingly $-\Lambda$ on the mirror
brane, although this value will be corrected by terms of order $\rho$
in the presence of matter on the branes),
so that $\rho_t = \Lambda+ \rho$, where $\rho$ now denotes the
energy density of normal matter or radiation on the brane, as opposed to
vacuum energy. One can choose $\Lambda_b$ to exactly cancel the
$\Lambda^2$ terms in eq.\ (\ref{BDLeq}), and furthermore fix the value of
$\Lambda$ in terms of $M$ and $M_P$:
\beqa
\label{Lambda1}
	\Lambda_b &=& - {\Lambda^2 \over 6M^3};\\
\label{Lambda2}
	\Lambda &=& \pm 6 {M^6\over M_P^2},
\eeqa
where $\pm$ refers to the two respective branes.  Condition (3) insures
the cancellation of the effect of $\Lambda_b$ by $\Lambda^2$ in (1),
whereas (4) adjusts the overall rate of expansion to agree with the
usual result.  
The new Friedmann equation then becomes the conventional one, 
plus a correction
which is quadratic in the density:
\beq
\label{oureq}
	{H^2 = \pm {\rpm\over 3 M_P^2}\left(1 \pm \rpm {M_P^2\over 12 M^6}
	\right) }
\eeq
We have distinguished the values of $\rho$ on the two branes by the subscript
to emphasize that they need not--in fact, cannot--be the same.
The brane with the positive solution has a rate of expansion
that is consistent with all current cosmological observations as long as the 
normal rate has been recovered by the epoch of nucleosynthesis,
which will be true if $0< \rho_{\sss+}\lsim 0.1\,(1\ {\rm MeV})^4 \ll \Lambda$.
One thus finds the constraint that 
\beq
\label{nucbound}
	M \gsim 10 {\rm\ TeV},
\eeq
which is not much more severe than other accelerator and astrophysical
limits that have recently been placed on the new gravity scale.  The other
brane must have $\rho_{\sss -} \leq 0$,
since otherwise $H^2 < 0$, which has no solution.

The condition (\ref{Lambda1}) is precisely what is needed to get a static
universe in the case of vanishing $\rho$: the negative cosmological
constant in the bulk cancels the positive $\Lambda^2$ from either brane.
The solutions to the Einstein equations in this case were recently studied
by Randall and Sundrum (RS) \cite{RS}, but for very different reasons: they
found that the weak scale hierarchy problem is naturally solved on one of
the branes, even if $M\sim M_P$, and $b_0 \sim 50 M_P^{-1}$.  This comes
about because the metric tensor has an exponential dependence on the
coordinate of the compact 5th dimension (see Figure 1).
Using the line element
\beq
\label{metric}
ds^2 = - n^2(t,y) dt^2 + a^2(t,y) \delta_{ij} dx^i dx^j + b(t,y)^2 dy^2,
\eeq
it is straightforward to verify the time-independent solution 
\beq
\label{RSsoln}
	a(y) = n(y) = a_0 e^{-k|y|}; \quad k = {b_0\Lambda\over 6M^3};
	\quad b(y) = b_0;
\eeq
One then observes that, even if all mass parameters in the Lagrangians for
matter on the branes are of the order $M_P$, the physical masses on the
brane at $y=1/2$ are suppressed by the factor $e^{-k/2}$, which can be of
order $M_W/M_P$ with only a moderate hierarchy between $b_0$ and $M_P^{-1}
\sim M^{-1}$.  Since $g_{\mu\nu}$ enters differently in the kinetic than
the mass terms for a scalar field, once the kinetic terms are canonically
normalized, masses get multiplied by $a(1/2)\sim e^{-k/2}$.  This idea
therefore appears to be a much more natural solution to the hierarchy
problem than the original proposal, which required $b_0 M$ to be of order
$(M_P/M)^{2/N}$, where $N$ is the number of extradimensions.

We now see that the static solution of RS is the
starting point for our idea, which is to recover the normal expansion of
the 3D universe by perturbing large, balancing cosmological constants in
the bulk and the branes by a small density of matter or radiation on the
branes.  Intuitively, it is clear that solutions with nonvanishing $\rho$
must exist, but we will now take some time to demonstrate this explicitly,
in the vicinity of our brane.  We were not able to find global solutions in
closed form once matter with an arbitrary equation of state $p =
\omega\rho$ was introduced.  However, we are really most interested in the
expansion rate on our own brane, so it suffices to solve the Einstein
equations in that region.  To simplify the appearance of the solutions, we
will translate the $y$ coordinate by $y\to y+1/2$, so that the brane which
we inhabit is located at $y=0$. 

We must solve the Einstein equations for the metric (\ref{metric}),
now allowing for time dependence in $a$, $b$ and $n$.  It is
always possible to chose a gauge so that $n(t,0)$ is constant at $y=0$,
without introducing $g_{05}$ elements in the metric.  We will make this
choice, and drop all terms involving $\dot n$ since they are not
relevant for the solution in the immediate vicinity of the brane.  With
this simplification, the 5D Einstein equations, $G_{\mu\nu} = 
M^{-3}T_{\mu\nu}$, become \cite{BDL}
\beqa
&&\fda \left( \fda+ \fdb \right) = \frac{n^2}{b^2}
\left(\fppa + \fpa \left( \fpa - \fpb \right) \right) + {1\over 3 M^3} 
T_{00}, \label{ein00}\\
&&\left(\fda\right)^2 +2\fdda + 2 \fdb \fda + \fddb = - {n^2\over 
a^2 M^3}T_{ii}
+ \frac{n^2}{b^2} \left\{ 2\fppa \right.\\
\nonumber\\
&&\left. +\fppn
\fpa\left(\fpa+2\fpn\right) - \fpb\left(\fpn+2\fpa\right)
\right\}
\label{einii} \\
&&\left(\fda\right)^2 + \fdda =  \frac{n^2}{b^2}
\fpa \left(\fpa+\fpn \right) - {n^2\over 3 b^2 M^3}T_{55}
\label{ein55}\\
&&\fpn \fda + \fpa \fdb - \frac{\dot{a}^{\prime}}{a}
= {1\over 3 M^3} T_{05} = 0
\label{ein05}
\end{eqnarray}
in the vicinity of
$y=0$.  Close to our brane, the nonzero elements of the 5D stress-energy 
tensor are
\beqa
	T_{00} &=& n^2 (\rho+\Lambda)\delta(by) + n^2 (\Lambda_b + V(b))\,; \nonumber\\
	T_{ii} &=& a^2 (p - \Lambda)\delta(by) - a^2 (\Lambda_b + V(b))\,; \nonumber\\
	T_{55} &=&  - b^2(\Lambda_b + V(b) + V'(b)/b)
\eeqa
where $\delta(by)= b^{-1}\delta(y)$ is the generally covariant form of
the delta function.
There are also source terms proportional to $b^{-1}\delta(y-1/2)$ at the
mirror brane, but these will not directly concern us in what follows.
The terms involving $V(b)$ would result if there is a potential that stabilizes
the compact dimension.  Their presence does not qualitatively change any
of our conclusions, so we set $V(b)$ to zero in what follows.

The generalization of the static solution (\ref{RSsoln}) can be
parametrized as
\beqa
\label{param}
a(t,y) &=& a_0(t) \exp(\sfrac12 A|y| + \sfrac12 A_2 y^2 + \cdots)\nonumber\\
b(t,y) &=& b_0 \exp(\sfrac12 B|y| + \sfrac12 B_2 y^2 + \cdots)\nonumber\\
n(t,y) &=& \exp(\sfrac12 \cN|y| + \sfrac12 \cN_2 y^2 + \cdots)
\eeqa
By our choice of gauge for time, there is no $n_0(t)$ function.  We
have not assumed separability of the solution here, since the coefficients
$A,B,\cN$, {\it etc.}, need not be static; however we will see that their
time-dependence arises entirely from that of $\rho$ and $p$.  The fact 
that $b_0$ is constant in time is not obvious, but will be proven to be consistent 
with  eqs.\ (\ref{ein00}-\ref{ein05}).

As in ref.\ \cite{BDL}, the linear-in-$|y|$ coefficients, $A$ and
$\cN$, are determined by the singular parts of eqs.\ (\ref{ein00}) and
(\ref{einii}), {\it i.e.}, those involving the delta functions and 
second spatial derivatives.  One finds that
\beqa
	A &=& -\sfrac13 b_0 M^{-3} (\rho + \Lambda)\,; \nonumber\\
	\cN &=&  b_0 M^{-3}(p + \sfrac23 \rho -\sfrac13\Lambda).
\eeqa
Therefore, to obtain solutions that are growing in the direction of the
mirror brane, as are needed to solve the hierarchy problem on our own,
{\it we would have to chose $\Lambda < 0$ here,} 
about which we shall say more below. The analogous coefficient $B$ is not
determined in this way because $b''$ appears nowhere in the Einstein
equations.  But it is constrained by eq.\ (\ref{ein05}).  Inserting the
ansatz (\ref{param}) in this equation, and taking $\omega = p/\rho$ to be
constant (which is a weak restriction since $p$ and $\rho$ refer only to
the matter and radiation), one can eventually show that
\beqa
\label{Beq}
B ={b_0\over M^{3}}\left(\rho + p -
\Lambda(1+\omega)\ln\left(1+{\rho\over\Lambda}\right)\right)
	+ {\cal O}(A_2,\cN_2).
\eeqa
and that it is consistent to take $\dot b_0 = 0$.  Thus the scale factor of
the compact dimension, although it expands inside the bulk, is strictly
constant on our brane.  Eq.\ (\ref{Beq}) is not a complete specification
for $B$ since $A_2$ and $\cN_2$ are not yet known, but in fact we will
never need $B$ for determining the Friedmann equation on our brane.

It remains to satisfy the nonsingular parts of the other Einstein equations, 
(\ref{ein00} -- \ref{ein55}), near $y=0$.  The knowledge of $A$ and $\cN$ is
all that is needed to specify eq.\ (\ref{ein55}) at $y=0$ because no
second derivatives appear.  One obtains
\beqa
\left(\fdao\right)^2 + \fddao &=& {1\over 36 M^6}
	(\Lambda+\rho)(2\Lambda - \rho-3p) + {\Lambda_b\over 3 M^3}
	\nonumber\\
	&=& {\rho - 3p \over 6 M_P^2} - {\rho(\rho+3p)\over 36 M^6},
\label{oureq2}
\eeqa
where the second equation follows from using our previous determination of
$\Lambda$ and $\Lambda_b$, eqs.\ (\ref{Lambda1}--\ref{Lambda2}).  The
leading term reproduces the usual prediction of general relativity, and the
second term corresponds to the quadratic correction in eq.\ (\ref{oureq}). 
Indeed, in light of the energy conservation law on the brane, 
$\dot\rho = -3H(\rho+p)$,
which is true regardless of the extra dimension \cite{BDL},
 (\ref{oureq}) is the only relation consistent with (\ref{oureq2}) 
when $\omega=p/\rho$ is assumed to be constant.

In contrast to the new Einstein equation (\ref{ein55}) associated with the
5th dimension, the $G_{00}$ and $G_{ii}$ equations (\ref{ein00}) and
(\ref{einii}) depend on the quadratic coefficients $A_2$ and $\cN_2$
at $y=0$, because of the presence of $a''$ and $n''$.  With two equations
in two unknowns, it is always possible to find values of $A_2$ and $\cN_2$
such that the resulting equations for $a_0(t)$ are consistent with 
(\ref{oureq}) and (\ref{oureq2}).  Therefore eqs.\ (\ref{ein00}) and
(\ref{einii}) add no new information on the brane, although they would be
necessary if one wanted to deduce the full $y$ dependence of the solutions
in the bulk.

In the above derivation, it was shown that the brane whose masses are
small by the RS mechanism must have $\Lambda <0$. Unfortunately, we
already saw in eq.\ (\ref{oureq}) that the brane with negative
$\Lambda$ must have an energy density $\rho_{\sss -} \le 0$, which is
not the case in our universe.  This would appear to be a serious
problem for the RS idea.  However, see the ``Note added,'' below.


The conditions (\ref{Lambda1}-\ref{Lambda2}) for the brane and bulk
cosmological constants look strange at first, so some words of motivation
are in order.  Although when $\rho=0$, $\Lambda_+ = -\Lambda_-$ on the
two branes, when $\rho\neq 0$,
a global solution to the Einstein equations is needed in order to derive 
the exact relation between $\Lambda_+$ and $\Lambda_-$ when $\rho\neq 0$,
since it involves all the coefficients of the expansion (\ref{param}) 
\cite{foot3}: %
\beq
\label{Lbrane}
\frac{b_+(\Lambda_++\rho_+)}{A/2}
=-
\frac{b_-(\Lambda_- +\rho_-)}{A/2+A_2 y_- +\ldots}
\eeq
In addition to this topological relation 
derived from the spacetime geometry, there is also a relation involving
$\Lambda_b$. We argue that the latter is a stringent consistency
condition similar to the global tadpole cancellation
in string theory (see for instance \cite{PW}):
\beq
	\label{tadpole}
{\sqrt{g} \over b} \Lambda_{|+}
+{\sqrt{g} \over b} \Lambda_{|-}
+\int_{-1/2}^{1/2} dy \sqrt{g}
\left( \Lambda_b  - \sfrac12 M^3 {\cal R}  \right)
=0
\eeq
{\it i.e.} the global effective cosmological constant must vanish.
In the solution of RS, this condition reduces to
$\Lambda_0^2 + 6M^3\Lambda_b=0$,
which is the relation they needed to obtain a global solution to Einstein
equations.

The condition (\ref{tadpole}) can be understood if the cosmological
constants are viewed as an effective description of the Ramond--Ramond
fields of the underlying string theory: for instance the value of the
$(p+1)$-form to which a $p$-brane is coupled is reinterpreted as a
cosmological constant on the $p$-brane. The condition (\ref{tadpole}) will
now be necessary to cancel the UV divergences of the string theory.  The
connection between the phenomenological scenario of RS and string 
theory has recently been examined by Verlinde \cite{Verlinde} and his
analysis concludes that the exponential dependence of the metric in the
compact direction is identified with the renormalization group scale when
using the AdS/CFT correspondence, which also corroborates the stringy
origin of the RS mechanism.

Since the normal expansion rate of the universe is only known to have
held between nucleosynthesis and the present epoch (as was stressed in
reference \cite{Joyce}) it would be interesting if the quadratic
corrections to the new Friedmann equation (\ref{oureq}) started to
become important above temperatures of several MeV.  In the most
natural version of the RS scenario, $M$ and $M_P$ are of the same
order, so the corrections become important only at the Planck scale.
However it is still a logical possibility to imagine that the
fundamental scale $M$ is much smaller than $M_P$.  In this case one
recovers the Arkani-Hamed {\it et al.} result that $M_P^2 = M^3 b_0$,
which combined with the gravitational tests that restrict $b_0 \lsim 1$
mm, gives the constraint $M>10^8$ GeV.  With such a large value of $M$,
departure from normal expansion occurs only above temperatures $T \gsim
1$ TeV, which is not far above the electroweak scale.  An intriguing
possibility would be to increase the rate of expansion during the
electroweak phase transition.  If this occurred, standard model
sphaleron interactions could easily be out of equilibrium in the broken
phase \cite{Joyce}, making electroweak baryogenesis more feasible.

We thank Emilian Dudas, Nemanja Kaloper, Guy Moore, Burt Ovrut, and Carlos
Savoy for useful discussions.  JC thanks the CEA Saclay theory group for
their kind hospitality.  As we were submitting this work, ref.\
\cite{CGKT} appeared, which reached conclusions similar to ours. 

{\it Note added:} After acceptance of this work, we discovered
\cite{CGS} that the problem of the wrong-sign expansion rate at the
second brane can be solved if the extra dimension is taken to be
noncompact, as suggested by ref.\ \cite{RL}.  Ref.\ \cite{CGS} shows
that by considering multiple intersecting branes, the whole
construction can be extended to any number of extra dimensions.



\begin{references}
\bibitem{foot3} For the compact slice of AdS-space corresponding
to the solution of RS, $b$ is constant
and the exponential in the expression of $a$ is limited to the linear
term; in this case, the relation (\ref{Lbrane}) simplifies and gives
$\Lambda_+ = - \Lambda_-$.
\end{references}
\end{document}